# An Extension for Combination of Duty Constraints in Role-Based Access Control

Ali Hosseini
ICT Group, E-Learning Center,
Iran University of Science and Technology
Tehran, Iran

Mohammad Abdollahi Azgomi (Corresponding Author)
School of Computer Engineering,
Iran University of Science and Technology
Tehran, Iran

*Abstract*—Among access control models, Role-Based Access Control (RBAC) is very useful and is used in many computer systems. Static Combination of Duty (SCD) and Dynamic Combination of Duty (DCD) constraints have been introduced recently for this model to handle dependent roles. These roles must be used together and can be considered as a contrary point of conflicting roles. In this paper, we propose several new types of SCD and DCD constraints. Also, we introduce strong dependent roles and define new groups of SCD constraints for these types of roles as SCD with common items and SCD with union items. In addition, we present an extension for SCD constraints in the presence of hierarchy.

I. INTRODUCTION

Role-Based Access Control (RBAC) model is accepted well for its numerous advantages such as policy neutrality and efficient access control management [1, 2, 3]. Furthermore, the concept of role is associated with the notion of functional role in an organization. Hence, RBAC models provide intuitive support for expressing organizational access control policies.

RBAC has been introduced by Ferraiolo and Kuhn in 1992 and then has been completed by Sandhu et al. in 1996 [4, 3]. In 2001, National Institute of Standards and Technology (NIST) provided a standard for RBAC model [1]. In this model, Static Separation of Duty (SSD) and Dynamic Separation of Duty (DSD) constraints have been defined. Also, SSD has been extended in the presence of hierarchy. SSD means that no user must be assigned to conflicting roles. DSD means that no user must activate conflicting roles within the session.

Ahn and Sandhu proposed Role-based Separation of duty Language (RSL99) in the context of RBAC model and then extended this language to Role-based constraints Language (RCL 2000) [5, 6]. Their language can specify Separation of Duty (SD) constraints for conflicting permissions and conflicting users.

Simon and Zurko classified completely SD constraints in the role-based environments [7]. They proposed object-based SD (i.e. no user may act upon a target that the user has previously acted upon), operational SD (i.e. no user may assume a set of roles that have the capability for a complete business job), history-based SD (i.e. no user is allowed to perform all the actions in a business task in the same target or collection of targets).

In [7], formal definitions of SD constraints are not provided. Gligor et al. formally defined these constraints and also presented them with more details [8]. For example, SSD constraint was defined in two forms of Strict SSD (SSSD) and one step SSSD (1sSSSD) [8]. SSSD means that conflicting roles are not permitted to perform more than one operation on objects. 1sSSSD means that each two distinct roles in a set of conflicting roles are not permitted to perform operations on the same object.

SD constraints are an important topic in RBAC model and many researchers have investigated their aspects and issues. SD constraints force conflicting items such as roles, permissions and so on to be used separately. But, we should not focus only on conflicting items; because, there are dependent items in our environments, which must be used together. In [9], Hosseini and Azgomi introduced Combination of Duty (CD) constraints which handle dependent roles. They proposed Static Combination of Duty (SCD) and Dynamic Combination of Duty (DCD). SCD means that a user must be assigned to dependent roles. DCD means that a user must activate dependent roles.

SCD and DCD constraints are not enough to support a wide range of dependent roles. Therefore, it is necessary to define more CD constraints. Also, CD constraints and dependent roles can be considered as a contrary point of SD constraints and conflicting roles, respectively. Hence, we can declare a new CD constraint corresponding to each SD constraint.

In this paper, we propose completely the following CD constraints:
- Two types of SCD,
- SCD with common items,
- SCD with union items,
- SCD, SCD with common items and SCD with union, items in the presence of hierarchy, and
- Five types of DCD.

Two types of SCD constraints distribute dependent roles between a set of users. SCD with common items constraints are a strict version of SCD types. Here, common items must be assigned to dependent roles. SCD with union items constraints are an intermediate version between SCD with common items and SCD constraints. In RBAC model, roles can inherit other roles. Therefore, we extend SCD constraints in this manner. Five types of DCD constraints distribute dependent roles between a set of sessions and a set of users. By using these new CD constraints, we can specify powerful, exact and flexible policies related to dependent roles.

The remainder of this paper is organized as follows. In Section II, the RBAC model and its required detail are mentioned. In section III, new CD constraints are formally defined. Finally, we conclude the paper and mention some future works in Section IV.

## II. THE NIST RBAC MODEL

In 2001, NIST provided a standard for RBAC model [1]. RBAC model is defined in terms of three components: core, hierarchy and constraint. Each component has some elements. Also, some functions are defined in NIST RBAC model.

In RBAC model, core includes six elements, called USERS, ROLES, OBS, OPS, PRMS and SESSIONS, which are the sets of users, roles, objects, operations, permissions and sessions, respectively [1]. User is defined as a human being and role is a job function in an organization. Permission is an approval to perform an operation (such as *read* or *write*) on one or more objects. Users and permissions are assigned to roles. UA and UP are sets of user-role and permission-role assignments, respectively. PRMS, UA and PA can be formally defined as follows:

$$PRMS = 2^{OPS \times OBS} \quad (1)$$

$$UA \subseteq USERS \times ROLES \quad (2)$$

$$PA \subseteq PRMS \times ROLES \quad (3)$$

We explain some required functions related to core as follows:
- *AssignedRoles* function returns a set of roles assigned to the user.
- *RolePrms* function returns a set of permissions assigned to the role.
- *RoleObs* function returns a set of objects assigned as permissions to the role.
- *RoleOps* function returns a set of operations assigned as permissions to the role.
- *RoleOpsOnOb* function returns a set of operations which can perform on an object by the role.

The above functions can be formally defined as follows:

$$AssignedRoles(u) = \{r: ROLES \mid (u, r) \in UA\} \quad (4)$$

$$RolePrms(r) = \{p: PRMS \mid (p, r) \in PA\} \quad (5)$$

$$RoleObs(r) = \{ob: OBS \mid \exists op \in OPS: ((op, ob), r) \in PA\} \quad (6)$$

$$RoleOps(r) = \{op: OPS \mid \exists ob \in OBS: ((op, ob), r) \in PA\} \quad (7)$$

$$RoleOpsOnOb(r, ob) = \{op: OPS \mid ((op, ob), r) \in PA\} \quad (8)$$

Each session is associated with a single user and each user is associated with one or more sessions. A user can activate a subset of roles assigned to him/her within a session. The following are some functions related to session:
- *UserSessions* function returns a set of sessions associated with the user.
- *SessionUser* function returns the user who is owner of the session.
- *SessionRoles* function returns a set of roles activated within the session.
- *ActivatedRoles* function returns a set of roles activated by the user.

The above functions can be formally defined as follows:

$$UserSessions(u) \rightarrow 2^{SESSIONS} \quad (9)$$

$$\forall u_1, u_2 \in USERS: UserSessions(u_1) \cap UserSessions(u_2) = \varnothing$$

$$SessionUser(s) = \{u: USERS \mid UserSession(u) \cap \{s\} = \{s\}\} \quad (10)$$

$$SessionRoles(s) \subseteq AssignedRoles(SessionUser(s)) \quad (11)$$

$$ActivatedRoles(u) = \bigcup_{s \in UserSessions(u)} SessionRoles(s) \quad (12)$$

A Role Hierarchy (RH) is a subset of ROLES×ROLES and defines a seniority relation between roles, whereby senior roles inherit permissions of their juniors, and junior roles inherit user membership of their seniors. This relation between roles $r_1$ and $r_2$ is denoted by $r_1 \succeq r_2$ or $(r_1, r_2) \in RH$, whereby $r_1$ and $r_1$ are senior and junior roles, respectively.

We use the term *authorized* to refer to both *assigned* and *inherited*. Now, we revise the hierarchal version of the above functions for core as follows:
- *AuthorizedRoles* function returns a set of roles, which are authorized for the user.
- *RoleHPrms* function returns a set of permissions, which are authorized for the role.
- *RoleHObs* function returns a set of objects, which are authorized as permissions for the role.
- *RoleHOps* function returns a set of operations, which are authorized as permissions for the role.
- *RoleHOpsOnHOb* function returns a set of operations, which can be performed on an object by the role in the presence of hierarchy.

The above functions can be formally defined as follows. In the following definitions, we have $\forall r \in ROLES: (r, r) \in RH$.

$$AuthorizedRoles(u) = \{r: ROLES \mid \exists r' \in ROLES: \quad (13)$$
$$(r', r) \in RH \land (u, r') \in UA\}$$

$$RoleHPrms(r) = \{p: PRMS \mid \exists r' \in ROLES: \quad (14)$$

$(r, r') \in$ RH $\wedge$ $(p, r') \in$ PA}

RoleHObs($r$)= {$ob$: OBS| $\exists r' \in$ ROLES, $op \in$ OPS:  (15)

$(r, r') \in$ RH $\wedge$ (($op, ob$), $r'$) $\in$ PA}

RoleHOps($r$)= {$op$: OPS| $\exists r' \in$ ROLES, $ob \in$ OBS:  (16)

$(r, r') \in$ RH $\wedge$ (($op, ob$), $r'$) $\in$ PA}

RoleHOpsOnHOb($r, ob$)={$op$: OPS| $\exists r' \in$ ROLES:  (17)

$(r, r') \in$ RH $\wedge$ (($op, ob$), $r'$) $\in$ PA}

Constrained RBAC includes three types of SD constraints as SSD, SSD in the presence of hierarchy and DSD.

**Definition 1.** Separation of Duty (SD) is defined as a pair ($rs, r_n$), which $rs$ is a set of conflicting roles and $r_n$ is a natural number greater than or equal to 2 and less than or equal to the cardinality of $rs$.

**Definition 2.** Static Separation of Duty (SSD) is a subset of SD, such that each user must be assigned to less than $r_n$ conflicting roles. SSD is formally defined as follows, where N is a set of natural numbers:

$$\text{SSD} \subseteq 2^{\text{ROLES}} \times \text{N} \quad (18)$$

$\forall (rs, r_n) \in$ SSD, $u \in$ USERS: |AssignedRoles($u$)$\bigcap rs$|<$r_n$

**Definition 3.** Static Separation of Duty in the presence of Hierarchy (SSDH) is a subset of SD, such that each user must be authorized for less than $r_n$ conflicting roles. SSDH is formally defined as follows:

$$\text{SSDH} \subseteq 2^{\text{ROLES}} \times \text{N} \quad (19)$$

$\forall (rs, r_n) \in$ SSDH, $u \in$ USERS: |AuthorizedRoles($u$)$\bigcap rs$|<$r_n$

**Definition 4.** Dynamic Separation of Duty (DSD) is a subset of SD, such that less than $r_n$ conflicting roles must be activated within the session. DSD is formally defined as follows:

$$\text{DSD} \subseteq 2^{\text{ROLES}} \times \text{N} \quad (20)$$

$\forall (rs, r_n) \in$ DSD, $s \in$ SESSIONS: |SessionRoles($s$)$\bigcap rs$|<$r_n$

It is worth to mention that the above formal definitions differ from that appeared in [1] for NIST RBAC model, but they have the same meanings. The above definitions are given in order to be compatible with the new constraints we present in the next section.

### III. DEFINITIONS OF EXTENDED COMBINATION OF DUTY CONSTRAINTS

In RBAC model and its extensions, which we introduced in section I, the conflicting roles have been investigated by numerous researchers. Hence, various SD constraints have been defined to handle them. Conflicting roles must be used separately. But we do not deal with only such roles. There is another group of roles, which are called dependent roles. These roles are as a contrary point of conflicting roles and must be used together. Because of dependency, if we assign them to distinct users, then we will waste much time to collaborate users. Therefore, using dependent roles separately is not desired and efficient and may be entirely impossible.

The importance of dependent roles undergoes a rising trend, because the new methods and tools are generated continuously, which automate the works. Hence, the required time for doing tasks are decreased. Therefore, more roles can be assigned to one person while it was not possible in the past. Organizations are very interested in this topic, because they are looking for the ways to reduce their expenses for better competition. It is obvious that they prefer to assign dependent roles instead of independent roles to one user. Therefore, some policies appear which are based on dependent roles and we need various constraints for specifying these policies. However, there are not enough constraints that are related to dependent roles.

In a previous paper [9], we introduced Combination of Duty (CD) constraints which handle dependent roles. CD is defined as a pair ($rs, r_n$) which $rs$ is a set of dependent roles and $r_n$ is a natural number greater than or equal to 1 and less than the cardinality of $rs$. Also, we defined Static Combination of Duty (SCD) and Dynamic Combination of Duty (DCD). We repeat the formal definitions of these constraints in the subsections A and E. Also, we define other types of CD constraints in the following subsections.

#### A. Various types of static combination of duty

In this subsection, first we present the formal definition of SCD constraint as appeared in [9]. Then, we propose two other types of SCD constraints. We rename the original SCD as SCD$_{\text{type I}}$ or shortly SCD$_\text{I}$ and call the new constraints as SCD$_{\text{II}}$ and SCD$_{\text{III}}$.

**Definition 5.** Type I of Static Combination of Duty (SCD$_\text{I}$) means that each user must be assigned to no or more than $r_n$ dependent roles. SCD$_\text{I}$ can be formally defined as follows:

$$\text{SCD}_\text{I} \subseteq 2^{\text{ROLES}} \times \text{N} \quad (21)$$

$\forall (rs, r_n) \in$ SCD$_\text{I}$, $u \in$ USERS:

|AssignedRoles($u$)$\bigcap rs$|=0 $\vee$ |AssignedRoles($u$)$\bigcap rs$|>$r_n$

**Example 1.** SCD$_\text{I}$ = {({$r_1, r_2, r_3, r_4$}, 2)},
AssignedRoles($u_1$)={$r_1, r_2, r_3$}, AssignedRoles($u_2$)={$r_5$}, AssignedRoles($u_3$)={$r_1$}

$u_1$ can satisfy the constraint because $u_1$ is assigned to more than two dependent roles. $u_2$ can satisfy the constraint, because $u_2$ is assigned to no dependent role. $u_3$ cannot satisfy the constraint, because $u_3$ is not assigned to enough dependent roles (i.e. more than two).

As mentioned earlier, dependent roles must be used together. SCD$_\text{I}$ focuses dependency on each user.

Therefore, he/she must be assigned to more than $r_n$ dependent roles. We define other versions, which distribute dependency between a set of users. Therefore, this set instead of each user must be assigned to more than $r_n$ dependent roles.

**Definition 6.** Type II of Static Combination of Duty (SCD$_{II}$) means that the user $u$ can be assigned to less than or equal to $r_n$ dependent roles if the following conditions are satisfied:
- There is the set $us$ of users assigned to less than or equal to $r_n$ dependent roles.
- $u$ and $us$ are assigned to more than $r_n$ dependent roles.

We explain the expression "less than or equal to $r_n$" of the first condition that $u$ and $us$ must need together to satisfy the dependency relation. If $us$ is assigned to more than $r_n$ dependent roles then $u$ will not have any effect in this satisfaction. SCD$_{II}$ can be formally defined as follows:

$$\text{SCD}_{II} \subseteq 2^{\text{ROLES}} \times N \quad (22)$$

$$\forall (rs, r_n) \in \text{SCD}_{II}:$$

$\forall u \in \text{USERS}$, such that $0 < |\text{AssignedRoles}(u) \cap rs| \leq r_n$:

$\exists us \subseteq \text{USERS}: |(\bigcup_{x \in us} \text{AssignedRoles}(x)) \cap rs| \leq r_n \;\wedge$

$|(\bigcup_{x \in us \cup \{u\}} \text{AssignedRoles}(x)) \cap rs| > r_n$

**Example 2.** SCD$_{II}$={({$r_1, r_2, r_3, r_4$}, 2)},
AssignedRoles($u_1$)={$r_1$}, AssignedRoles($u_2$)={$r_2, r_3$},
AssignedRoles($u_3$)={$r_2$}, AssignedRoles($u_4$)={$r_3$},
AssignedRoles($u_5$)={$r_1, r_2, r_3$}.

$u_5$ is assigned to more than two dependent roles. Hence, this user does not need other users to satisfy SCD$_{II}$. $u_1$ and $u_2$ are not assigned to enough dependent roles. It is true because {$u_1, u_2$} is assigned to enough dependent roles. Also, $u_1$, $u_3$ and $u_4$ are not assigned to enough dependent roles. It is true because {$u_1, u_3, u_4$} is assigned to enough dependent roles.

As observed, $u_1$ is a common user of {$u_1, u_2$} and {$u_1, u_3, u_4$}. Therefore, we define SCD$_{III}$ that forces the sets to be distinct:

**Definition 7.** Type III of Static Combination of Duty (SCD$_{III}$) means that the user $u$ can be assigned to less than or equal to $r_n$ dependent roles if the following conditions are satisfied:
- There is the set $us$ of users assigned to less than or equal to $r_n$ dependent roles.
- $u$ and $us$ are assigned to more than $r_n$ dependent roles.
- $u$ and $us$ do not have this relation with other users.

SCD$_{III}$ can be formally defined as follows:

$$\text{SCD}_{III} \subseteq 2^{\text{ROLES}} \times N \quad (23)$$

$$\forall (rs, r_n) \in \text{SCD}_{III}:$$

$\exists us_1, us_2, \ldots, us_n \subseteq \text{USERS}: \bigcup_{i=1}^{n} us_i = \text{USERS} \;\wedge\; (\forall us_i,$

$us_j$ such that $i \neq j: us_i \cap us_j = \emptyset) \;\wedge\; (\forall us_i: |(\bigcup_{x \in us_i} \text{Assigned}$

$\text{Roles}(x)) \cap rs| = 0 \;\vee\; ((\bigcup_{x \in us_i} \text{AssignedRoles}(x)) \cap rs| > r_n \;\wedge$

$(\forall uss \subset us_i$ such that $|uss| = |us_i| - 1: |(\bigcup_{x \in uss} \text{AssignedRoles}$

$(x)) \cap rs| \leq r_n)))$

**Example 3.** SCD$_{III}$={({$r_1, r_2, r_3, r_4$}, 2)}

**Step 1:**

AssignedRoles($u_1$)={$r_1$}, AssignedRoles($u_2$)={$r_2, r_3$},
AssignedRoles($u_3$)={$r_2$}, AssignedRoles($u_4$)={$r_3$}.

$us_1$={$u_1, u_2$} is assigned to more than two dependent roles and can satisfy SCD$_{III}$. Users of this set such as $u_1$ cannot cooperate with other users such as $u_3$ and $u_4$. Hence, we cannot consider $us_2$={$u_1, u_3, u_4$}. Also, {$u_3, u_4$} is not assigned to enough dependent roles. We have two ways to satisfy SCD$_{III}$. In the first way, $u_3$ or $u_4$ is assigned to $r_1$ or $r_4$ (i.e. AssignedRoles($u_3$)={$r_1, r_2$}). Therefore, $us_2$={$u_3, u_4$} can satisfy SCD$_{III}$. In the second way, another user such $u_5$ is assigned to $r_1$ or $r_4$ (i.e. AssignedRoles($u_5$)={$r_1$}). Therefore, $us_2$={$u_3, u_4, u_5$} can satisfy SCD$_{III}$.

**Step 2:**

AssignedRoles($u_1$)={$r_1$}, AssignedRoles($u_2$)={$r_2$},
AssignedRoles($u_3$)={$r_3, r_4$}.

$us_1$={$u_1, u_2, u_3$} cannot satisfy SCD$_{III}$, because a subset of $us_1$ such as {$u_2, u_3$} is assigned to more than $r_n$ dependent roles. Also, we cannot consider $us_1$={$u_1, u_3$} and $us_2$={$u_2, u_3$}, because these two sets have a common user (i.e. $u_3$). If $u_3$ is de-assigned from $r_4$, then $us_1$={$u_1, u_2, u_3$} can satisfy SCD$_{III}$.

*B. Static Combination of Duty with common objects, operations and permissions*

In this subsection, we define new constraints as SCD with common items, which are strict versions of SCD types. These constraints force *strong dependency* between roles. Here, common items such as objects, operations and permissions must be assigned to dependent roles. We can define four kinds of these constraints for each type of SCD as follows:
- SCD with common objects.
- SCD with common operations.
- SCD with common objects and operations.
- SCD with common permissions.

Common items are specified in two ways. In the first way, a set of common items is determined. We use *obs*, *ops* and *prms*, which are sets of common objects, operations and permissions, respectively. These sets cannot be empty and must have some members. In the second way, the number of common items is determined.

We use $ob_n$, $op_n$ and $prm_n$, which are the number of common objects, operations and permissions, respectively. These parameters cannot be zero and must be a natural number. In the formal definitions, we use them as ($obs$||$ob_n$), ($ops$||$op_n$) and ($prms$||$prm_n$). The symbol "||" between two expressions means that only one of the two expressions must be considered. Also, we say that "||" may be used for more times in a definition or paragraph. For all "||", only the left expressions or only the right expressions must be considered.

We define $SCD_I$ with common items. Definitions of other types are similar to them.

**Definition 8.** Type I of Static Combination of Duty with Common Object ($SCDCOB_I$) means that each user must be assigned to no or more than $r_n$ dependent roles. Also, in the second state, the intersection of objects assigned as permissions to these roles must include the set $obs$ or have more than or equal to $ob_n$ members. $SCDCOB_I$ can be formally defined as follows:

$$SCDCOB_I \subseteq 2^{ROLES} \times N \times (2^{OBS}||N) \quad (24)$$

$\forall\ (rs, r_n, obs||ob_n) \in SCDCOB_I,\ u \in USERS$:

$|rss|=0\ \lor\ (|rss|>r_n\ \land\ (obs \subseteq rssobs\ ||\ ob_n \leq |rssobs|))$

, such that $rss=AssignedRoles(u) \bigcap rs\ \land$

$$rssobs = \bigcap_{r \in rss} RoleObs(r).$$

**Definition 9.** Type I of Static Combination of Duty with Common Operations ($SCDCOP_I$) means that each user must be assigned to no or more than $r_n$ dependent roles. Also, in the second state, the intersection of operations assigned as permissions to these roles must include the set $ops$ or have more than or equal to $op_n$ members. $SCDCOP_I$ can be formally defined as follows:

$$SCDCOP_I \subseteq 2^{ROLES} \times N \times (2^{OPS}||N) \quad (25)$$

$\forall\ (rs, r_n, ops||op_n) \in SCDCOP_I,\ u \in USERS$:

$|rss|=0\ \lor\ (|rss|>r_n\ \land\ (ops \subseteq rssops\ ||\ op_n \leq |rssops|))$

, such that $rss=AssignedRoles(u) \bigcap rs\ \land$

$$rssops = \bigcap_{r \in rss} RoleOps(r).$$

We join SCDCOB and SCDCOP together to define a stricter version than them. In this version, common operations must perform on common objects.

**Definition 10.** Type I of Static Combination of Duty with Common Objects and Operations ($SCDCOB\text{-}OP_I$) means that each user must be assigned to no or more than $r_n$ dependent roles. Also, in the second state, the intersection of objects assigned as permissions to these roles must include the set $obs$ or have more than or equal to $ob_n$ members. In addition, the intersection of operations, which can be performed on each common object by these roles must include the set $ops$ or have more than or equal

to $op_n$ members. $SCDCOB\text{-}OP_I$ can be formally defined as follows:

$$SCDCOB\text{-}OP_I \subseteq 2^{ROLES} \times N \times (2^{OBS}||N) \times (2^{OPS}||N) \quad (26)$$

$\forall\ (rs, r_n, obs||ob_n, ops||op_n) \in SCDCOB\text{-}OP_I,\ u \in USERS$:

$|rss|=0\ \lor\ (|rss|>r_n\ \land\ (obs \subseteq rssobs\ ||\ ob_n \leq |rssobs|)\ \land$

$(\forall\ ob \in obs:\ ops \subseteq rssopson(ob)\ ||\ op_n \leq |rssopson(ob)|))$

, such that $rss=AssignedRoles(u) \bigcap rs\ \land$

$$rssobs = \bigcap_{r \in rss} RoleObs(r)\ \land$$

$$rssopson(ob) = \bigcap_{r \in rss} RoleOpsOnOb(r, ob)).$$

SCDCOB-OP forces to perform all common operations on each common object. But it may be needed to perform some operations on $object_1$, some operations on $object_2$ and so on. This demand can be fulfilled by SCD with common permissions. We mention that the permission is an approval to perform an operation on an object.

**Definition 11.** Static Combination of Duty with Common Permissions ($SCDCPRMS_I$) means that each user must be assigned to no or more than $r_n$ dependent roles. Also, in the second state, the intersection of permissions assigned to these roles must include the set $prms$ or have more than or equal to $prm_n$ members. $SCDCPRMS_I$ can be formally defined as follows:

$$SCDCPRMS_I \subseteq 2^{ROLES} \times N \times (2^{PRMS}||N) \quad (27)$$

$\forall\ (rs, r_n, prms||prm_n) \in SCDCPRMS_I,\ u \in USERS$:

$|rss|=0\ \lor\ (|rss|>r_n\ \land\ (prms \subseteq rssprms\ ||\ prm_n \leq |rssprms|))$

, such that $rss=AssignedRoles(u) \bigcap rs\ \land$

$$rssprms = \bigcap_{r \in rss} RolePrms(r).$$

**Example 4.** There are the following specifications.

$RolePrms(r_1)=\{(ob_1, op_1), (ob_1, op_2), (ob_2, op_1), (ob_2, op_2)\}$

$RolePrms(r_2)=\{(ob_1, op_1), (ob_1, op_2), (ob_2, op_1), (ob_2, op_2), (ob_3, op_3)\}$,

$RolePrms(r_3)=\{(ob_1, op_1), (ob_3, op_3)\}$,

$RolePrms(r_4)=\{(ob_1, op_1), (ob_2, op_2), (ob_4, op_4)\}$,

$RolePrms(r_5)=\{(ob_3, op_3), (ob_4, op_4)\}$,

$RolePrms(r_6)=\{(ob_1, op_1), (ob_1, op_2), (ob_1, op_3)\ (ob_2, op_1), (ob_2, op_2), (ob_2, op_3)\}$,

As a result, we have:

$RoleObs(r_1)=\{ob_1, ob_2\}$, $\quad RoleObs(r_2)=\{ob_1, ob_2, ob_3\}$,

$RoleObs(r_3)=\{ob_1, ob_3\}$, $\quad RoleObs(r_4)=\{ob_1, ob_2, ob_4\}$,

$RoleObs(r_5)=\{ob_3, ob_4\}$, $\quad RoleObs(r_6)=\{ob_1, ob_2\}$,

$RoleOps(r_1)=\{op_1, op_2\}$, $\quad RoleOps(r_2)=\{op_1, op_2, op_3\}$,

$RoleOps(r_3)=\{op_1, op_3\}$, $\quad RoleOps(r_4)=\{op_1, op_2, op_4\}$,

RoleOps($r_5$)={$op_3$, $op_4$},   RoleOps($r_6$)={$op_1$, $op_2$, $op_3$},
RoleOpsOnOb($r_1$, $ob_1$)={$op_1$, $op_2$},
RoleOpsOnOb($r_1$, $ob_2$)={$op_1$, $op_2$},
RoleOpsOnOb($r_2$, $ob_1$)={$op_1$, $op_2$},
RoleOpsOnOb($r_2$, $ob_2$)={$op_1$, $op_2$},
RoleOpsOnOb($r_2$, $ob_3$)={$op_3$},
RoleOpsOnOb($r_3$, $ob_1$)={$op_1$},
RoleOpsOnOb($r_3$, $ob_3$)={$op_3$},
RoleOpsOnOb($r_4$, $ob_1$)={$op_1$},
RoleOpsOnOb($r_4$, $ob_2$)={$op_2$},
RoleOpsOnOb($r_4$, $ob_4$)={$op_4$},
RoleOpsOnOb($r_5$, $ob_3$)={$op_3$},
RoleOpsOnOb($r_5$, $ob_4$)={$op_4$},
RoleOpsOnOb($r_6$, $ob_1$)={$op_1$, $op_2$, $op_3$},
RoleOpsOnOb($r_6$, $ob_2$)={$op_1$, $op_2$, $op_3$}.

**Step 1:** SCDCOB$_I$={({$r_1$, $r_2$, $r_3$, $r_4$}, 2, {$ob_1$, $ob_2$})}.

AssignedRoles($u_1$)={$r_1$, $r_2$, $r_3$} cannot satisfy the constraint, because the intersection of objects assigned to these roles (i.e. {$ob_1$}) does not include $obs$ (i.e. $obs$={$ob_1$, $ob_2$}).

AssignedRoles($u_2$)={$r_1$, $r_2$, $r_4$} can satisfy the constraint, because the intersection of objects assigned to these roles (i.e. {$ob_1$, $ob_2$}) includes $obs$.

**Step 2:** SCDCOB$_I$={({$r_1$, $r_3$, $r_5$}, 1, 1)}.

AssignedRoles($u_3$)={$r_1$, $r_3$} can satisfy the constraint, because the intersection of objects assigned to these roles (i.e. {$ob_1$}) has $ob_n$ member (i.e. $ob_n$=1).

AssignedRoles($u_4$)={$r_3$, $r_5$} can satisfy the constraint, because the intersection of objects assigned to these roles (i.e. {$ob_3$}) has $ob_n$ member.

**Step 3:** SCDCOP$_I$={({$r_1$, $r_2$, $r_3$, $r_4$}, 2, {$op_1$, $op_2$})}.

AssignedRoles($u_1$)={$r_1$, $r_2$, $r_3$} cannot satisfy the constraint, because the intersection of operations assigned to these roles (i.e. {$op_1$}) does not include $ops$ (i.e. $ops$={$op_1$, $op_2$}).

AssignedRoles($u_2$)={$r_1$, $r_2$, $r_4$} can satisfy the constraint, because the intersection of operations assigned to these roles (i.e. {$op_1$, $op_2$}) includes $ops$.

**Step 4:** SCDCOB-OP$_I$={({$r_1$, $r_2$, $r_3$, $r_4$}, 2, {$ob_1$, $ob_2$}, {$op_1$, $op_2$})}.

AssignedRoles($u_2$)={$r_1$, $r_2$, $r_4$} cannot satisfy the constraint, because although the intersection of objects assigned to these roles (i.e. {$ob_1$, $ob_2$}) includes $obs$ (i.e. $obs$={$ob_1$, $ob_2$}), but the intersection of operations which can be performed on each common object by these roles (i.e. {$op_1$} on $ob_1$ and {$op_2$} on $ob_2$) does not include $ops$ (i.e. $ops$={$op_1$, $op_2$}).

AssignedRoles($u_5$)={$r_1$, $r_6$} can satisfy the constraint, because the intersection of objects assigned to these roles (i.e. {$ob_1$, $ob_2$}) includes $obs$. Also, the intersection of operations, which can be performed on each common object by these roles (i.e. {$op_1$, $op_2$} on $ob_1$ and {$op_1$, $op_2$} on $ob_2$) includes $ops$.

**Step 5:** SCDCPRMS$_I$={({$r_1$, $r_2$, $r_3$, $r_4$}, 2, {($ob_1$, $op_1$), ($ob_2$, $op_2$)})}.

AssignedRoles($u_2$)={$r_1$, $r_2$, $r_4$} can satisfy the constraint, because the intersection of permissions assigned to these roles (i.e. {($ob_1$, $op_1$), ($ob_2$, $op_2$)}) includes $prms$ (i.e. $prms$={($ob_1$, $op_1$), ($ob_2$, $op_2$)}).

*C. Static Combination of Duty with union objects, operations and permissions*

SCD with common items constraints focus on each dependent role and force common items to be assigned to it. In this subsection, we define new constraints as SCD with union items, which focus on a set of dependent roles and forces the items to be assigned to this set. These constraints can be considered as an intermediate version between SCD with common items and SCD constraints. Similar to SCD with common items, there are four kinds of SCD with union items constraints for each type of SCD as follows:

- SCD with union objects.
- SCD with union operations.
- SCD with union objects and operations.
- SCD with union permissions.

We define SCD$_I$ with union objects and operations. Definitions of other kinds and types are similar to it.

**Definition 12.** Type I of Static Combination of Duty with Union Objects and Operations (SCDUOB-OP$_I$) means that each user must be assigned to no or more than $r_n$ dependent roles. Also, in the second state, union of objects assigned as permissions to these roles must include the set $obs$ or have more than or equal to $ob_n$ members. In addition, union of operations, which can be performed on each union object by these roles must include the set $ops$ or have more than or equal to $op_n$ members. SCDUOB-OP$_I$ can be formally defined as follows:

$$\text{SCDUOB-OP}_I \subseteq 2^{\text{ROLES}} \times N \times (2^{\text{OBS}} \| N) \times (2^{\text{OPS}} \| N) \quad (28)$$

$\forall (rs, r_n, obs\|ob_n, ops\|op_n) \in$ SCDUOB-OP$_I$, $u \in$ USERS:

$|rss|=0 \lor (|rss|>r_n \land (obs \subseteq rssobs \| ob_n \leq |rssobs|) \land$

$(\forall ob \in obs: ops \subseteq rssopson(ob) \| op_n \leq |rssopson(ob)|))$

, such that $rss$=AssignedRoles($u$)$\bigcap rs \land$

$$rssobs = \bigcup_{r \in rss} \text{RoleObs}(r) \land$$

$$rssopson(ob) = \bigcup_{r \in rss} \text{RoleOpsOnOb}(r, ob)).$$

**Example 5**: There are the following specifications.

RolePrms($r_1$)={($ob_1$, $op_1$), ($ob_2$, $op_1$)},
RolePrms($r_2$)={($ob_1$, $op_2$)},   RolePrms($r_3$)={($ob_3$, $op_4$)},
RolePrms($r_4$)={($ob_1$, $op_3$), ($ob_2$, $op_2$)},

As a result, we have:

RoleObs($r_1$)={$ob_1$, $ob_2$},    RoleObs($r_2$)={$ob_1$},
RoleObs($r_3$)={$ob_3$},    RoleObs($r_4$)={$ob_1$, $ob_2$},
RoleOps($r_1$)={$op_1$},    RoleOps($r_2$)={$op_2$},
RoleOps($r_3$)={$op_4$},    RoleOps($r_4$)={$op_2$, $op_3$},
RoleOpsOnOb($r_1$, $ob_1$)={$op_1$},
RoleOpsOnOb($r_1$, $ob_2$)={$op_1$},
RoleOpsOnOb($r_2$, $ob_1$)={$op_2$},
RoleOpsOnOb($r_3$, $ob_3$)={$op_4$},
RoleOpsOnOb($r_4$, $ob_1$)={$op_3$},
RoleOpsOnOb($r_4$, $ob_2$)={$op_2$}.

**Step 1:** SCDUOB$_I$={({$r_1$, $r_2$, $r_3$, $r_4$}, 2, {$ob_1$, $ob_2$})}.

AssignedRoles($u_1$)={$r_1$, $r_2$, $r_3$} can satisfy the constraint, because the union of objects assigned to these roles (i.e. {$ob_1$, $ob_2$, $ob_3$}) includes *obs* (i.e. *obs*={$ob_1$, $ob_2$}).

**Step 2:** SCDUOP$_I$={({$r_1$, $r_2$, $r_3$, $r_4$}, 2, {$op_1$, $op_2$})},

AssignedRoles($u_1$)={$r_1$, $r_2$, $r_3$} can satisfy the constraint, because the union of operations assigned to these roles (i.e. {$op_1$, $op_2$, $op_4$}) includes *ops* (i.e. *ops*={$op_1$, $op_2$}).

**Step 3:** SCDUOB-OP$_I$={({$r_1$, $r_2$, $r_3$, $r_4$}, 2, {$ob_1$, $ob_2$}, {$op_1$, $op_2$})}.

AssignedRoles($u_2$)={$r_1$, $r_2$, $r_3$} cannot satisfy the constraint, because although the union of objects assigned to these roles (i.e. {$ob_1$, $ob_2$, $ob_3$}) includes *obs* (i.e. *obs*={$ob_1$, $ob_2$}) but the union of operations which can perform on each union object by these roles (i.e. {$op_1$, $op_2$} on $ob_1$ and {$op_1$} on $ob_2$) does not include *ops* (i.e. *ops*={$op_1$, $op_2$}).

AssignedRoles($u_5$)={$r_1$, $r_2$, $r_4$} can satisfy the constraint, because the union of objects assigned to these roles (i.e. {$ob_1$, $ob_2$}) includes *obs*. Also, the union of operations which can perform on each union object by these roles (i.e. {$op_1$, $op_2$, $op_3$} on $ob_1$ and {$op_1$, $op_2$} on $ob_2$) includes *ops*.

**Step 4:** SCDUPRMS$_I$={({$r_1$, $r_2$, $r_3$, $r_4$}, 2, {($ob_1$, $op_1$), ($ob_2$, $op_2$)})}.

AssignedRoles($u_2$)={$r_1$, $r_2$, $r_3$} cannot satisfy the constraint, because the union of permissions assigned to these roles (i.e. {($ob_1$, $op_1$), ($ob_1$, $op_2$), ($ob_2$, $op_1$), ($ob_3$, $op_4$)}) does not include *prms* (i.e. *prms*={($ob_1$, $op_1$), ($ob_2$, $op_2$)}).

AssignedRoles($u_2$)={$r_1$, $r_2$, $r_4$} can satisfy the constraint, because the union of permissions assigned to these roles (i.e. {($ob_1$, $op_1$), ($ob_1$, $op_2$), ($ob_1$, $op_3$), ($ob_2$, $op_1$), ($ob_2$, $op_2$)}) includes *prms*.

### D. Static Combination of Duty in the presence of hierarchy

In this subsection, we revise the previous constraints in the presence of hierarchy, which we already explained in section II. We define hierarchal versions of SCD$_I$, SCD$_{II}$, SCDCOB-OP$_I$ and SCDUOB-OP$_I$. Definitions of the other constraints are similar to them.

**Definition 13.** Type I of Static Combination of Duty in the presence of Hierarchy (SCDH$_I$) means that each user must be authorized for no or more than $r_n$ dependent roles. SCDH$_I$ can be formally defined as follows:

$$SCDH_I \subseteq 2^{ROLES} \times N \qquad (29)$$

$$\forall (r_n, n) \in SCDH_I, u \in USERS:$$

$$|AuthorizedRoles(u) \cap rs|=0 \lor |AuthorizedRoles(u) \cap rs|>r_n$$

**Definition 14.** Type II of Static Combination of Duty in the presence of Hierarchy (SCDH$_{II}$) means that the user $u$ can be authorized for less than or equal to $r_n$ dependent roles if the following conditions are satisfied.
- There is the set *us* of users authorized for less than or equal to $r_n$ dependent roles.
- $u$ and *us* are authorized for more than $r_n$ dependent roles.

SCDH$_{II}$ can be formally defined as follows:

$$SCDH_{II} \subseteq 2^{ROLES} \times N \qquad (30)$$

$$\forall (rs, r_n) \in SCDH_{II}:$$

$$\forall u \in USERS, \text{such that } 0<|AuthorizedRoles(u) \cap rs| \leq r_n:$$

$$\exists us \subseteq USERS: |(\bigcup_{x \in us} AuthorizedRoles(x)) \cap rs| \leq r_n \land$$

$$|(\bigcup_{x \in us \cup \{u\}} AuthorizedRoles(x)) \cap rs|>r_n$$

**Definition 15.** Type I of Static Combination of Duty in the presence of Hierarchy with Common Objects and Operations (SCDHCOB-OP$_I$) means that each user must be authorized for no or more than $r_n$ dependent roles. Also, in the second state, the intersection of objects authorized as permissions for these roles must include the set *obs* or have more than or equal to $ob_n$ members. In addition, the intersection of operations which can be performed on each object by these roles in the presence of hierarchy must include the set *ops* or have more than or equal to $op_n$ members. SCDHCOB-OP$_I$ can be formally defined as follows:

$$SCDHCOB\text{-}OP_I \subseteq 2^{ROLES} \times N \times (2^{OBS}\|N) \times (2^{OPS}\|N) \qquad (31)$$

$$\forall (rs, r_n, obs\|ob_n, ops\|op_n) \in SCDHCOB\text{-}OP_I, u \in USERS:$$

$$|rss|=0 \lor (|rss|>r_n \land (obs \subseteq rsshobs \| ob_n \leq |rsshobs|) \land$$

$$(\forall ob \in obs: ops \subseteq rsshopsonh(ob) \| op_n \leq |rsshopsonh(ob)|))$$

, such that $rss = AuthorizedRoles(u) \cap rs \land$

$$rsshobs = \bigcap_{r \in rss} RoleHObs(r) \land$$

$$rsshopsonh(ob) = \bigcap_{r \in rss} RoleHOpsOnHOb(r, ob)).$$

**Definition 16.** Type I of Static Combination of Duty in the presence of Hierarchy with Union Objects and Operations (SCDHUOB-OP$_I$) means that each user must be authorized for no or more than $r_n$ dependent roles. Also, in the second state, the union of objects authorized as permissions for these roles must include the set *obs* or have more than or equal to $ob_n$ members. In addition, the union of operations which can be performed on each object by these roles in the presence of hierarchy must include the set *ops* or have more than or equal to $op_n$ members. SCDHUOB-OP$_I$ can be formally defined as follows:

$$\text{SCDHUOB-OP}_I \subseteq 2^{ROLES} \times N \times (2^{OBS} \| N) \times (2^{OPS} \| N) \quad (32)$$

$\forall (rs, r_n, obs\|ob_n, ops\|op_n) \in \text{SCDHUOB-OP}_I, u \in \text{USERS}:$

$|rss|=0 \lor (|rss|>r_n \land (obs \subseteq rsshobs \| ob_n \leq |rsshobs|) \land$

$(\forall ob \in obs: ops \subseteq rsshopsonhob \| op_n \leq |rsshopsonh(ob)|))$

,such that $rss = \text{AuthorizedRoles}(u) \bigcap rs \land$

$$rsshobs = \bigcup_{r \in rss} \text{RoleHObs}(r) \land$$

$$rsshopsonh(ob) = \bigcup_{r \in rss} \text{RoleHOpsOnHOb}(r, ob)).$$

**Example 6**. There are following specifications.
RH={$(r_3, r_2)$}.
RolePrms($r_1$)={$(ob_1, op_1), (ob_2, op_1)$},
RolePrms($r_2$)={$(ob_1, op_1), (ob_2, op_2)$},
RolePrms($r_3$)={$(ob_1, op_4)$},  RolePrms($r_4$)={$(ob_1, op_2)$}.

As a result, we have:
RoleObs($r_1$)={$ob_1, ob_2$},  RoleObs($r_2$)={$ob_1, ob_2$},
RoleObs($r_3$)={$ob_1$},  RoleObs($r_4$)={$ob_1$},
RoleHObs($r_3$)={$ob_1, ob_2$},
RoleOps($r_1$)={$op_1$},  RoleOps($r_2$)={$op_1, op_2$},
RoleOps($r_3$)={$op_4$},  RoleOps($r_4$)={$op_2$},
RoleHOps($r_3$)={$op_1, op_2, op_4$},
RoleOpsOnOb($r_1, ob_1$)={$op_1$},
RoleOpsOnOb($r_1, ob_2$)={$op_1$},
RoleOpsOnOb($r_2, ob_1$)={$op_1$},
RoleOpsOnOb($r_2, ob_2$)={$op_2$},
RoleOpsOnOb($r_3, ob_1$)={$op_4$},
RoleOpsOnOb($r_3, ob_2$)={ },
RoleOpsOnOb($r_4, ob_1$)={$op_2$},
RoleHOpsOnHOb($r_3, ob_1$)={$op_1, op_4$},
RoleHOpsOnHOb($r_3, ob_2$)={$op_2$}.

The hierarchal versions of functions are equal to original functions for roles, which are not senior of other roles such as $r_1$, $r_2$ and $r_4$. For example, RoleHObs($r_1$)=RoleObs($r_1$) and RoleHOpsOnHOb($r_1$)=RoleObsOnOb($r_1$).

**Step 1:**
**Part 1:** SCD$_I$={({$r_1, r_2, r_3, r_4$}, 2)}.

AssignedRoles($u_1$)={$r_1, r_3$} cannot satisfy the constraint, because this user is not assigned to enough dependent roles (i.e. more than 2).

**Part 2:** SCDH$_I$={({$r_1, r_2, r_3, r_4$}, 2)}.

AssignedRoles($u_1$)={$r_1, r_3$} can satisfy the constraint, because this user is authorized for more than two dependent roles (i.e. {$r_1, r_2, r_3$}).

**Step 2:**
**Part 1:** SCDCOB$_I$={({$r_1, r_2, r_3, r_4$}, 2, {$ob_1, ob_2$})}.

AssignedRoles($u_1$)={$r_1, r_3$} cannot satisfy the constraint, because this user is not assigned to enough dependent roles. Also, the intersection of objects assigned to these roles (i.e. {$ob_1$}) does not include *obs* (i.e. *obs*={$ob_1, ob_2$}).

**Part 2:** SCDHCOB$_I$={({$r_1, r_2, r_3, r_4$}, 2, {$ob_1, ob_2$})}.

AssignedRoles($u_1$)={$r_1, r_3$} can satisfy the constraint, because this user is authorized for more than two dependent roles and the intersection of objects authorized for these roles (i.e. {$ob_1, ob_2$}) includes *obs*.

**Step 3:**
**Part 1:** SCDUOB-OP$_I$={({$r_1, r_2, r_3, r_4$}, 2, {$ob_1, ob_2$}, {$op_1, op_2$})}.

AssignedRoles($u_2$)={$r_1, r_3, r_4$} cannot satisfy the constraint, because the union of operations which can be performed on each object by these roles (i.e. {$op_1, op_2, op_4$} on $ob_1$ and {$op_1$} on $ob_2$) does not include *ops* (i.e. *ops*={$op_1, op_2$}).

**Part 2:** SCDHUOB-OP$_I$={({$r_1, r_2, r_3, r_4$}, 2, {$ob_1, ob_2$}, {$op_1, op_2$})}.

AssignedRoles($u_2$)={$r_1, r_3, r_4$} can satisfy the constraint, because the union of objects assigned to these roles (i.e. {$ob_1, ob_2$}) includes *obs* (i.e. *obs*={$ob_1, ob_2$}) and the union of the operations, which can be performed on each object by these roles in the presence of hierarchy (i.e. {$op_1, op_2, op_4$} on $ob_1$ and {$op_1, op_2$} on $ob_2$) includes *ops*.

*E. Various types of dynamic combination of duty*

In this subsection, first we present the formal definition of DCD constraint which was already defined in [9]. This constraint is in the base of each session. Therefore, we rename it as DCD$_{\text{S-type I}}$ or shortly DCD$_{\text{S-I}}$. We define another version which is based on each user. This version is called as DCD$_{\text{U-I}}$. Then, we propose two other types of each constraint as DCD$_{\text{S-II}}$, DCD$_{\text{S-III}}$, DCD$_{\text{U-II}}$ and DCD$_{\text{U-III}}$.

**Definition 17.** Type S-I of Dynamic Combination of Duty (DCD$_{\text{S-I}}$) means that no or more than $r_n$ dependent roles must be activated within the session. DCD$_{\text{S-I}}$ can be formally defined as follows:

$$\text{DCD}_{\text{S-I}} \subseteq 2^{ROLES} \times N \quad (33)$$

$\forall (rs, r_n) \in \text{DCD}_{\text{S-I}}, s \in \text{SESSIONS}:$

$|\text{SessionRoles}(s) \bigcap rs|=0 \lor |\text{SessionRoles}(s) \bigcap rs|>r_n$

**Example 7**. DCD$_{\text{S-I}}$={({$r_1, r_2, r_3, r_4$}, 2)}.

SessionRoles($s_1$)={$r_1$, $r_2$, $r_3$}, SessionRoles($s_2$)={$r_5$}, SessionRoles($s_3$)={$r_1$}.

$s_1$ can satisfy the constraint, because more than two dependent roles are activated within it. $s_2$ can satisfy the constraint, because none of the dependent roles are activated within it. $s_3$ cannot satisfy the constraint, because enough dependent roles are not activated within it (i.e. more than two).

**Definition 18.** Type U-I of Dynamic Combination of Duty (DCD$_{U-I}$) means that no or more than $r_n$ dependent roles must be activated by the user. DCD$_{U-I}$ can be formally defined as follows:

$$\text{DCD}_{U-I} \subseteq 2^{\text{ROLES}} \times N \quad (34)$$

$$\forall (rs, r_n) \in \text{DCD}_{U-I}, u \in \text{USERS}:$$

$$|\text{ActivatedRoles}(u) \cap rs|=0 \vee |\text{ActivatedRoles}(u) \cap rs|>r_n$$

**Example 8.** DCD$_{U-I}$={({$r_1$, $r_2$, $r_3$, $r_4$}, 2)},
UserSessions($u_1$)={$s_1$, $s_2$, $s_3$, $s_4$}.
SessionRoles($s_1$)={$r_1$}, SessionRoles($s_2$)={$r_2$},
SessionRoles($s_3$)={ }, SessionRoles($s_4$)={$r_2$, $r_3$}.

As a result, we have:
ActivatedRoles($u_1$)={$r_1$, $r_2$, $r_3$}.

Enough dependent roles are not activated within each session of $u_1$. But, enough dependent roles are activated by $u_1$. Therefore, DCD$_{U-I}$ is satisfied. As observed, DCD$_{U-I}$ distribute dependency between the sessions of each user.

DCD$_{S-I}$ and DCD$_{U-I}$ focus dependency on each session and user, respectively. Hence, more than $r_n$ dependent roles must be activated within/by each session/user. We define other versions which distribute dependency between a set of sessions and a set of users, respectively. Therefore, more than $r_n$ dependent roles must be activated within/by these sets instead of each session and user.

**Definition 19.** Type S-II of Dynamic Combination of Duty (DCD$_{S-II}$) means that less than or equal to $r_n$ dependent roles can be activated within the session $s$ if following conditions are satisfied.
- There is the set $ss$ of sessions that less than or equal to $r_n$ dependent roles are activated within it.
- More than $r_n$ dependent roles are activated within $s$ and $ss$.

SCD$_{S-II}$ can be formally defined as follows:

$$\text{DCD}_{S-II} \subseteq 2^{\text{ROLES}} \times N \quad (35)$$

$$\forall (rs, r_n) \in \text{DCD}_{S-II}:$$

$$\forall s \in \text{SESSIONS}, \text{such that } 0<|\text{SessionRoles}(s) \cap rs| \leq r_n:$$

$$\exists ss \subseteq \text{SESSIONS}: |(\bigcup_{x \in ss} \text{SessionRoles}(x)) \cap rs| \leq r_n \wedge$$

$$|(\bigcup_{x \in ss \cup \{s\}} \text{SessionRoles}(x)) \cap rs| > r_n$$

**Example 9.** DCD$_{S-II}$={({$r_1$, $r_2$, $r_3$, $r_4$}, 2)},
UserSessions($u_1$)={$s_1$, $s_2$, $s_3$, $s_4$}, UserSessions($u_2$)={$s_5$, $s_6$},
UserSessions($u_3$)={$s_7$, $s_8$, $s_9$},
AssignedRoles($s_1$)={$r_1$}, AssignedRoles($s_5$)={$r_2$, $r_3$},
AssignedRoles($s_2$)={$r_2$}, AssignedRoles($s_9$)={$r_3$},
AssignedRoles($s_7$)={$r_1$, $r_2$, $r_3$}.

More than two dependent roles are activated with $s_7$. Hence, this session does not need other sessions to satisfy DCD$_{S-II}$. Enough dependent roles are not activated within $s_1$ and $s_5$. It is true, because enough dependent roles are activated within {$s_1$, $s_5$}. Also, enough dependent roles are not activated within $s_1$, $s_2$ and $s_9$. It is true, because enough dependent roles are activated within {$s_1$, $s_2$, $s_9$}.

**Definition 20.** Type U-II of Dynamic Combination of Duty (DCD$_{U-II}$) means that the user $u$ can activate less than or equal to $r_n$ dependent roles if the following conditions are satisfied.
- There is the set $us$ of users which activates less than or equal to $r_n$ dependent roles.
- $u$ and $us$ activate more than $r_n$ dependent roles.

DCD$_{U-II}$ can be formally defined as follows:

$$\text{DCD}_{U-II} \subseteq 2^{\text{ROLES}} \times N \quad (36)$$

$$\forall (rs, r_n) \in \text{DCD}_{U-II}:$$

$$\forall u \in \text{USERS}, \text{such that } 0<|\text{ActivatedRoles}(u) \cap rs| \leq r_n:$$

$$\exists us \subseteq \text{USERS}: |(\bigcup_{x \in us} \text{ActivatedRoles}(x)) \cap rs| \leq r_n \wedge$$

$$|(\bigcup_{x \in us \cup \{u\}} \text{ActivatedRoles}(x)) \cap rs| > r_n$$

**Example 10.** DCD$_{U-II}$={({$r_1$, $r_2$, $r_3$, $r_4$}, 2)},
ActivatedRoles($u_1$)={$r_1$, $r_2$}, ActivatedRoles($u_2$)={$r_2$, $r_3$},
ActivatedRoles($u_3$)={$r_1$, $r_2$, $r_3$}, ActivatedRoles($u_4$)={$r_3$}.

$u_3$ activates more than two dependent roles. Hence, this user does not need other users to satisfy DCD$_{U-II}$. $u_1$ and $u_2$ do not activate enough dependent roles. It is true, because {$u_1$, $u_2$} activates enough dependent roles. Also, $u_1$ and $u_4$ do not activate enough dependent roles. It is true, because {$u_1$, $u_4$} activates enough dependent roles.

As showed in example 9, $s_1$ is a common session of {$s_1$, $s_5$} and {$s_1$, $s_2$, $s_9$}. Also, in example 10, $u_1$ is a common user of {$u_1$, $u_2$} and {$u_1$, $u_4$}. Therefore, we define DCD$_{S-III}$ and DCD$_{U-III}$ which force the sets to be distinct.

**Definition 21.** Type S-III of Dynamic Combination of Duty (DCD$_{S-III}$) means that less than or equal to $r_n$ dependent roles can be activated within the session $s$ if the following conditions are true.
- There is the set $ss$ of sessions that less than or equal to $r_n$ dependent roles are activated within it.
- More than $r_n$ dependent roles are activated within $s$ and $ss$.
- $s$ and $ss$ do not have this relation with other sessions.

$DCD_{S-III}$ can be formally defined as follows:

$$DCD_{S-III} \subseteq 2^{ROLES} \times N \quad (37)$$

$$\forall (rs, r_n) \in DCD_{S-III}:$$

$$\exists ss_1, ss_2, \ldots, ss_n \subseteq SESSIONS: \bigcup_{i=1}^{n} ss_i = SESSIONS \land$$

$$(\forall us_i, us_j \text{ such that } i \neq j: us_i \cap us_j = \emptyset) \land$$

$$(\forall ss_i: |(\bigcup_{x \in ss_i} SessionRoles(x)) \cap rs| = 0 \lor ((\bigcup_{x \in ss_i} SessionRole(x)) \cap rs| > r_n \land (\forall sss \subset us_i \text{ such that } |sss| = |ss_i| - 1:$$

$$|(\bigcup_{x \in sss} SessionRoles(x)) \cap rs| \leq r_n)))$$

**Example 11.** $DCD_{S-III} = \{(\{r_1, r_2, r_3, r_4\}, 2)\}$,
AssignedRoles($s_1$)={$r_1$}, AssignedRoles($s_5$)={$r_2, r_3$},
AssignedRoles($s_2$)={$r_2$}, AssignedRoles($s_9$)={$r_3$},
AssignedRoles($s_3$)={$r_1$}.

$ss_1=\{s_1, s_5\}$, $ss_2=\{s_2, s_3, s_9\}$.

$ss_1$ and $ss_2$ do not have common members. Also, more than two dependent roles are activated within them. Therefore, they can satisfy $DCD_{S-III}$.

**Definition 22.** Type U-III of Dynamic Combination of Duty ($DCD_{U-III}$) means that the user $u$ can activate less than or equal to $r_n$ dependent roles if following conditions are satisfied.
- There is the set $us$ of users which activates less than or equal to $r_n$ dependent roles.
- $u$ and $us$ activate more than $r_n$ dependent roles.
- $u$ and $us$ do not have this relation with other users.

$DCD_{U-III}$ can be formally defined as follows:

$$DCD_{U-III} \subseteq 2^{ROLES} \times N \quad (38)$$

$$\forall (rs, r_n) \in DCD_{U-III}:$$

$$\exists us_1, us_2, \ldots, us_n \subseteq USERS: \bigcup_{i=1}^{n} us_i = USERS \land (\forall us_i,$$

$us_j$ such that $i \neq j: us_i \cap us_j = \emptyset) \land (\forall us_i: |(\bigcup_{x \in us_i} Activated$

$Roles(x)) \cap rs| = 0 \lor ((\bigcup_{x \in us_i} ActivatedRoles(x)) \cap rs| > r_n \land$

$(\forall uss \subset us_i \text{ such that } |uss| = |us_i| - 1: |\bigcup_{x \in uss} ActivatedRoles$

$(x)) \cap rs| \leq r_n)))$

**Example 12.** $DCD_{U-III} = \{(\{r_1, r_2, r_3, r_4\}, 2)\}$,
ActivatedRoles($u_1$)={$r_1, r_2$}, ActivatedRoles($u_2$)={$r_2, r_3$},
ActivatedRoles($u_3$)={$r_1, r_2$}, ActivatedRoles($u_4$)={$r_3$}.

$us_1=\{u_1, u_2\}$, $us_2=\{u_3, u_4\}$.

$us_1$ and $us_2$ do not have common members. Also, they activate more than two dependent roles. Therefore, they can satisfy $DCD_{U-III}$.

## IV. CONCLUSIONS

In this paper, we proposed various types of combination of duty constraints for handling different aspects of dependent roles. Type I of SCD and DCD focus dependency on each user and session. Type II and III of SCD and DCD constraints distribute dependency between a set of users and a set of sessions. SCD with common items and SCD with union items are strict versions of SCD. The former forces common items to be assigned to each dependent role and the latter forces union items to be assigned to a set of dependent roles. Hierarchal version of SCD constraints considers seniority relation between roles as well as dependency relation. Also, we explain that the usage of dependent roles undergoes a rising trend. We want to do more research about dependent roles in the future works.